\documentstyle[prl,aps,twocolumn,epsf]{revtex}
\catcode`\@=11

\def\maketitle2{\par % Uses \twocolumn[\@maketitle2].
\begingroup
\let\cite\@bylinecite
\def\thefootnote{\fnsymbol{footnote}}%
\twocolumn[\@maketitle2\vskip2pc]%
\thispagestyle{plain}\@thanks
\endgroup
\def\thefootnote{\arabic{footnote}}%
\setcounter{footnote}{0}%
\let\maketitle2\relax \let\@maketitle2\relax
\let\@thanks\relax \let\@authoraddress\relax \let\@title\relax
\let\@date\relax \let\thanks\relax \let\@abstract\relax 
\let\@pacs\relax}

\def\abstract#1{\gdef\@abstract{{\par % Store abstract text. 
\bgroup
\ifdim\prevdepth=-1000pt \prevdepth0pt\fi
\hsize\columnwidth
\dimen0=-\prevdepth \advance\dimen0 by17.5pt \nointerlineskip
\small\vrule width 0pt height\dimen0 \relax}{~~}#1\egroup}}

\def\pacs#1{\gdef\@pacs{{\par % Store PACS numbers as \@pacs.
\bgroup
\hsize\columnwidth \parindent0pt
\ifdim\prevdepth=-1000pt \prevdepth0pt\fi
\dimen0=-\prevdepth \advance\dimen0 by20pt\nointerlineskip
\egroup} PACS numbers:~#1}}

\def\@maketitle2{% Puts \@abstract and \@pacs in a {list}.
\@preprint
\@title
\ifdim\prevdepth=-1000pt \prevdepth0pt\fi
\@authoraddress
\@date
\begin{list}{}{\leftmargin=0.10753\textwidth \rightmargin=\leftmargin
\itemsep=1pc\partopsep=-1pc}
\item\@abstract
\item\@pacs
\end{list}
}

\catcode`\@=12

\makeatletter

\def\compoundrel#1\over#2{\mathpalette\compoundreL{{#1}\over{#2}}}
\def\compoundreL#1#2{\compoundREL#1#2}
\def\compoundREL#1#2\over#3{\mathrel
  {\vcenter{\hbox{$\m@th\buildrel{#1#2}\over{#1#3}$}}}}
\makeatother

\begin{document}

\draft

\title{Decoherence and the rate of entropy production in chaotic quantum 
systems}

\author{Diana Monteoliva and 
Juan Pablo Paz \thanks{email addresses: monteoli@df.uba.ar, paz@df.uba.ar}}

\address{Departamento de F\'{\i}sica ``J.J. Giambiagi'', 
FCEN, UBA, Pabell\'on 1, Ciudad Universitaria, 1428 Buenos Aires, Argentina}

%\maketitle

\abstract
{We show that for an open quantum system which is classically chaotic 
(a quartic double well with harmonic driving coupled to a sea of harmonic
oscillators) the rate of entropy production has, as a function of time, two
relevant regimes: For short times it is proportional to the diffusion
coefficient (fixed by the system--environment coupling strength).  
For longer times (but before equilibration) there is a regime where the 
entropy production rate is fixed by the Lyapunov exponent. The nature of
the transition time between both regimes is investigated.}

\date{\today}

%%%%%%%%%%%%%%%%%%%%%%%%%%%%%%%%%%%%%%%%%%%%%%%%%%%%%%%%%%%
%
%		Main text
%
%%%%%%%%%%%%%%%%%%%%%%%%%%%%%%%%%%%%%%%%%%%%%%%%%%%%%%%%%%%

\maketitle2
\narrowtext

Environment induced decoherence has been identified in recent years as 
one of the main ingredients in the transition from quantum to 
classical behavior. Classicality emerges as a consequence of the coupling
of quantum systems to an environment which, in effect, dynamically enforces
super-selection rules by precluding the stable existence of the majority
of states in the Hilbert space of the system. The physics of decoherence
has been thoroughly studied during the last few years both from the 
theoretical \cite{decoherencereview} and also from the experimental point of 
view \cite{decoherenceexp}. 
As part of these studies it has been recognized that the decoherence 
process has rather unique properties for systems whose classical analogues  
are chaotic\cite{zp94}. In fact, for classically chaotic systems 
decoherence has two very important features: First, decoherence restores 
the validity of the correspondence principle that for isolated chaotic 
systems could be violated in an absurdly short time-scale (see the Hyperion
paradox in \cite{zp95}). Thus, environment induced
decoherence ensures that classical and quantum expectation values agree 
for a long range of times pushing towards infinity the ``breakdown' time 
which would otherwise be present and would depend logarithmically on $\hbar$.
This important feature 
was first conjectured in \cite{zp94} and later studied in more detail in 
\cite{zp95,hsz98,brumer}.

The second salient feature of decoherence for classically chaotic systems 
concerns the entropy production rate. In fact, in \cite{zp94} it was also 
conjectured that, when decoherence is effective, there is a robust range of 
parameters for which the entropy production rate becomes independent of the 
strength of the coupling to the environment. In such regime, the rate is
equal to the sum of positive Lyapunov exponents (the conjecture was 
modified in \cite{pata} where it was noticed that, in general, one 
should use appropriately averaged Lyapunov exponents). 

In this letter we present the first conclusive numerical evidence supporting 
the above conjecture for open quantum systems evolving continuously in time
(in the last few years the status of the conjecture was analyzed in part
for some open quantum maps \cite{hu95,sarkar,pata}).  
Our results show that Lyapunov exponent determines the value of the entropy 
production rate after an initial transient. We also clarify 
other important issues: First we show that during the
initial transient the entropy production rate is 
proportional to the system--environment coupling strength. 
Second we analyze the transition between the two 
regimes. Our numerical results are consistent with a linear dependence
of the transition time $t_c$ on the entropy of the initial state 
and with a logarithmic dependence on the coupling 
strength. Again, remarkably enough, 
our results validate once more the grossly oversimplified but very 
intuitive picture that was presented in \cite{zp94}. 

The system we analyze is a quantum particle moving in a quartic 
double well potential under the action of an harmonic driving force. 
The Hamiltonian is $H= p^2/2+V(x,t)$ with the potential  
$V(x,t)= - B x^2 + C x^4/2+ E x \cos(\omega t)$. This driven system has
been extensively studied \cite{doublewell} and it is well known that for a 
wide range of parameters it has chaotic (mixed) behavior. In our studies we 
chose several sets of parameters in such a way that the stroboscopic phase 
space portrait is like  
the one shown in Figure 1 where one clearly sees the coexistence of islands 
of stability and a chaotic sea. The coupling of this system to  
an environment is modeled via the simplest (ohmic, high temperature)  
quantum Brownian motion model. In this case, the evolution of the reduced 
density matrix of the system follows a local master equation \cite{QBM}, 
which gives rise to the following equation for the Wigner function of the 
system:
\begin{eqnarray}
\dot W=\left\{H,W\right\}_{PB}&+&\sum_{n\le 1}{(-1)^n\hbar^{2n}
\over{2^{2n}(2n+1)!}}\partial_x^{(2n+1)}V
\partial_p^{(2n+1)} W \nonumber\\
&+&2\gamma
\partial_p\left(p W\right)+D\partial_{pp}^2W\label{wignereq}
\end{eqnarray}
The physical effects included in this equation are well understood. The
first term in the right hand side is the Poisson bracket that induce 
classical evolution (Liouville flow). The second term carries the 
quantum corrections. 
The last two terms bring the influence of the environment 
generating, respectively, friction and diffusion. The regime of interest for 
us is the one where the friction term can be neglected ($\gamma\rightarrow 0$,
and energy is approximately conserved) while the diffusive term, 
that produces decoherence is non--negligible ($D=2m\gamma k_BT=$ constant).

\begin{figure}
\epsfxsize=8.6cm
\epsfbox{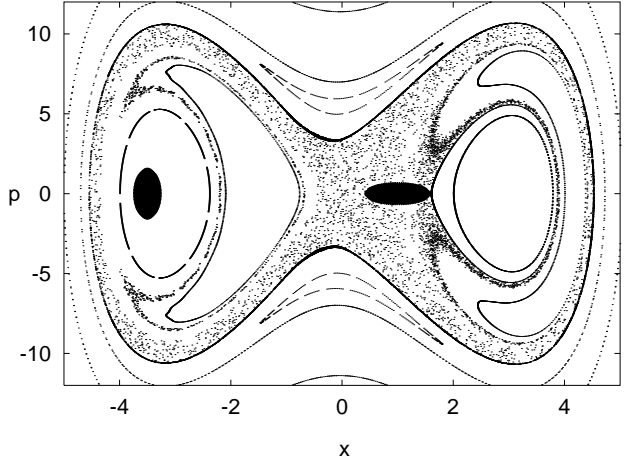}
\vspace {0.5cm}
\caption{Stroboscopic phase space portrait of the driven double well with 
parameters $B=10, C=0.5$, $E=1, \omega=5.35$. The black ellipses 
(one in a regular island and the other in the chaotic sea) show 
two of the initial states we considered (these are pure states and 
the boundary of the ellipses, whose area is $\pi\ln(20)\hbar\approx 9.4\hbar$, 
corresponds to points where the Wigner function decreases to $1/20$ of its 
peak value). Throughout our studies we used $\hbar =0.1$. }
\label{figure1}
\end{figure}

Let us first qualitatively describe the most important effects we expect to  
test in our numerical study. We consider initial states 
with a smooth Gaussian Wigner function centered either in the chaotic sea 
or in a regular island as shown in Figure 1 (these two cases 
lead to drastically different results). Let us suppose first that 
the diffusion term in (\ref{wignereq}) is absent (i.e., the state  
evolves according to Schr\"odinger equation). For a smooth initial 
state the dominant term in (\ref{wignereq}) is the Poisson bracket. As  
Wigner function initially evolves following nonlinear classical trajectories 
it looses its Gaussian shape and develops tendrils while folding (this happens 
exponentially fast if the initial state is in the chaotic sea). As a 
consequence, gradients of $W$ increase and quantum corrections 
in (\ref{wignereq}) become important, 
inducing oscillations in the Wigner function. The decoherence producing term 
in (\ref{wignereq}) can be qualitatively understood as being responsible 
of two interrelated effects: On the one hand, the diffusion term tends 
to smooth out the Wigner function naturally reducing gradients. 
This washes out the oscillations in the Wigner function 
suppressing quantum interference. 
The time-scale characterizing the dissapearence of 
the fringes can be estimated using previous  
results \cite{phz92}: Fringes with a characteristic wave-vector (along 
the $p$--axis of phase space) $k_p$ decay exponentially with a rate 
given by $\Gamma_D=D k_p^2$. Noting that a wave-packet spread over a 
distance $\Delta_x$ with two coherently interfering pieces generate
fringes with  $k_p=\Delta_x/\hbar$ one concludes that the decoherence 
rate is $\Gamma_D=D\Delta_x^2/\hbar^2$. This rate, depends linearly 
on the diffusion constant, and is one of the factors that contribute to 
the rate of entropy increase during the first phase of the evolution.  

There is a second related aspect of decoherence which is drastically 
different for regular and chaotic systems. Thus, appart from suppressing
the fringes, the diffusion term also tends to spread the regions where
the Wigner function is possitive, contributing in this way to 
the entropy growth. But, as discussed in \cite{zp94,pata}, the rate of 
entropy production distinguishes 
regular and chaotic cases. For regular states, decoherence 
should produce entropy at a rate which depends on the diffusion
constant $D$. However, for chaotic states the rate should become independent
of $D$ and should be fixed by the Lyapunov exponent. The origin 
of this $D$--independent phase can be understood using a simple minded 
argument (presented first in \cite{zp94} and later discussed in a more 
elaborate way in \cite{pata}): Chaotic dynamics tends to contract 
the Wigner function along some directions in phase space competing against 
diffusion. These two effects may balance
each other giving rise to a critical width below which Wigner function 
cannot contract. This local width should be approximately 
$\sigma_c^2=2D/\lambda$ (being $\lambda$ the  local Lyapunov exponent). 
Once this critical size has been reached, the 
contraction stops along the stable direction while the expansion continues
along the unstable one. Therefore, the area covered by the 
Wigner function grows exponentially in time and, as a consequence, entropy
grows linearly with a rate fixed by the Lyapunov exponents. 

Here, we will present solid numerical evidence supporting the 
existence of this $D$--independent phase. 
For simplicity, instead of looking at the von Neuman entropy 
${\cal H}_{VN}=-Tr\left(\rho_r\log\rho_r\right)$ we examine the linear 
entropy, defined as ${\cal H}=-\log(Tr\left(\rho_r^2\right))$, 
which is a good measure of the degree of mixing of the 
system and sets a lower bound on ${\cal H}_{VN}$. The above argument 
concerning the role of the critical width $\sigma_c$ may appear as  
too simple but captures the essential aspects of 
the dynamical process. Indeed, the master equation can be used to show that  
\begin{equation}
\dot{\cal H}=2 D 
\langle (\partial_p W)^2\rangle/\langle W^2\rangle\label{hdot}
\end{equation}
where the bracket denotes an integral over phase space. 
The right hand side of this equation is proportional to the mean square
wave--number computed with the Fourier transform of the Wigner function. 
This implies that the entropy production rate is closely related 
to the phase space structure present in the Wigner distribution. Thus, the 
$D$--independent phase begins at the time when the mean square 
wave--length along the momentum axis scales with diffusion as 
$\sqrt D$ (as $\sigma_c$ does). This behavior cancels the diffusion 
dependence of $\dot{\cal H}$ which becomes entirely determined by
the dynamics. 

Appart from analyzing the $D$--independent phase 
of entropy production we will analyze the nature of the the transition 
time between the diffusion dominated and the chaotic regime. This time 
$t_c$  can also be naively estimated along the lines of the previous
argument: The time for which the spread of the Wigner function 
approaches the critical one is  
$t_c\approx \lambda^{-1}\log(\sigma_p(0)/\sigma_c)$. Thus, one expects 
$t_c$ to depend logarithmically on the diffusion constant and
on the initial spread of the Wigner function (for Gaussian initial states 
the spread depends  exponentially on the initial entropy, therefore $t_c$ 
should vary as a linear function of the initial entropy). Our numerical 
work is devoted at testing these intuitive ideas. 

We solved numerically the master equation using two different methods (that 
give similar results). First, we computed the Wigner function 
numerically solving equation (\ref{wignereq}) using a third order 
pseudo--spectral method \cite{feit}. For this, 
we used a phase space grid of variable size and adaptive 
time-steps (a similar method was used in \cite{hsz98}). Second we 
obtained approximate solutions to (\ref{wignereq}) using a  
different strategy: We computed the Floquet states of the original 
(isolated) driven system and obtained the master equation in the 
Floquet basis \cite{masterfloquet}. This equation has a rather 
simple form when averaged over one driving period. In such case,
it can be numerically solved by usual means (the limiting factor 
is the number of Floquet states that are required). In 
what follows we present our results for the entropy behavior (plots are
obtained using the first of the two numerical methods, more details and 
other results will be presented elsewhere \cite{diana2}).  

The drastic difference between the behavior of the entropy production 
rate for regular and chaotic initial conditions is clearly displayed in 
Figure 2. Regular initial conditions produce entropy at a rate linearly 
dependent on the diffusion coefficient. This is precisely what we expect
when the entropy production is due to: (i) the destruction of interference 
fringes, (ii) the slow increase in the area covered by the possitive Wigner
function. The oscillations evident in Figure 2 have some distinguishing 
features: they have the frequency of the driving force and both their 
amplitude and their phase are $D$--independent (this is true for the 
regular initial state and for the initial transient of the chaotic state). 
These oscillations can be shown to be related both to the change in 
orientation of the fringes (decoherence is more effective if fringes are 
aligned along the $p$ axis) and to the change 
in spread of the Wigner function in the momentum direction induced by the 
dynamics. When the initial state is centered in the 
chaotic sea, the initial transient is followed by a regime 
where the rate $\dot{\cal H}$ becomes {\it independent} of the value 
of the diffusion constant (if $D$ is not too small, see below). 
Moreover, the numerical value of the rate 
oscillates around the average local Lyapunov exponent computed by 
averaging over an ensemble of classical trajectories 
weighted by the initial Wigner function (In \cite{pata} the use of 
a more elaborate averaging scheme was suggested, we will compare these 
approaches elsewhere \cite{diana2}). 

\begin{figure}
\epsfxsize=8.6cm
\epsfbox{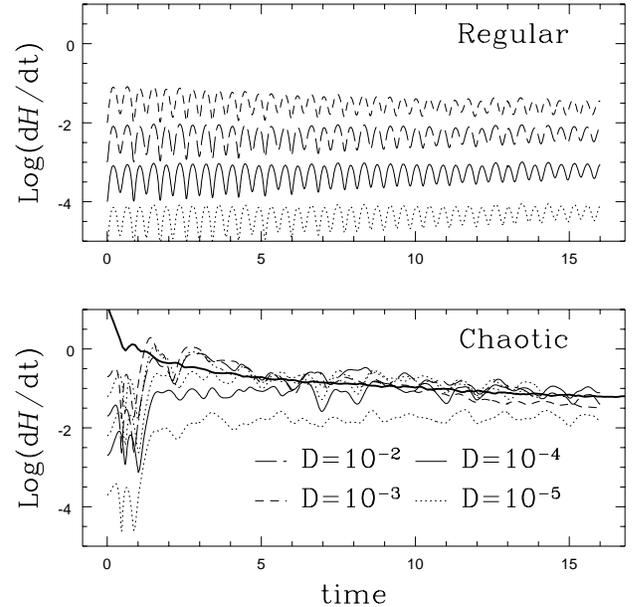}
\caption{Entropy production rate vs time (in units of the driving period). 
The bold curve is the (time dependent) Lyapunov exponent. 
The linear dependence of the rate on $D$ appears 
in the graph at the top (regular initial state) and
during the initial transient in the lower plot. In this case 
(initial state in the chaotic sea) the rate becomes independent on 
diffusion and is equal to the Lyapunov exponent (if $D$ the is not too 
small, see text).}

\label{figure2}
\end{figure}

It is remarkable that for long times the 
entropy production rate is indeed fixed just by the dynamics becoming 
independent of $D$ (after all, the entropy production is itself a consequence
of the coupling to the environment but the value of the rate becomes
independent of it!). The result presented in Figure 2 was shown to be 
robust under changes of initial conditions and other parameters 
characterizing the classical dynamics. 
There are two limitations for the above results to be obtained. On the one
hand the diffusion constant cannot be too strong: In that case the 
system heats up too fast and entropy saturates, making the numerical 
simulation unreliable. On the other hand, diffusion cannot be too small 
either: If that is the case decoherence could become too weak and the 
interference fringes could persist over many oscillations (the minimal value 
of $D$ required for efficient decoherence could be estimated as follows: 
if the Wigner function is coherently spread over a region of size
$\Delta_x\approx 10$, we would need a diffusion constant larger than
$D_{min}\approx\hbar^2/ \Delta_x^2\approx 10^{-4}$ for the environment to 
be able to wash out the smallest fringes in one driving period. In fact, 
Figure 2 shows that when $D=10^{-5}$ the entropy production rate is 
one order of magnitude smaller than the one corresponding to $D=10^{-4}$.

\begin{figure}
\epsfxsize=8.6cm
\epsfbox{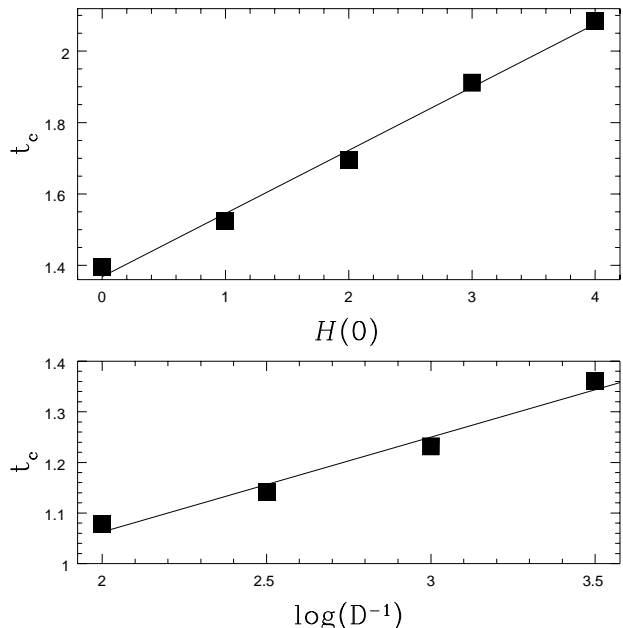}
\caption{The transition time between the diffusion dominated regime and
the one where the entropy production rate is set by the Lyapunov exponent is
shown to depend linearly on entropy (top) and logarithmically on the diffusion 
constant (bottom). Numerical results were obtained using the parameters:
$B=10, C=0.5, E=10, \omega=6.16$, $D=10^{-3}$ (top), 
${\cal H}(0)=0$ (bottom)}

\label{figure3}
\end{figure}

It is interesting to remark that while the simple picture presented in 
\cite{zp94} is in good qualitative agreement with our results, 
the arguments presented in that paper are too simple to include some 
important effects we found. In particular, the 
oscillatory nature of the rate was completely 
overlooked in \cite{zp94}. However, having said this, it is still possible
to test some simple results obtained in \cite{zp94} for the 
transition time between the diffusion dominated 
regime and the one where the rate is fixed by the Lyapunov exponent. 
For this purpose we performed two related studies. First we analyzed the 
dependence of the transition time $t_c$ on the diffusion constant. Of course,  
there is some ambiguity in the definition of the transition time 
because of the oscillatory nature of the rate. Here, we defined $t_c$ as the
time for which the rate reaches some value after the initial transient
(fortunately the value of the rate before and after the transient differ 
by two orders of magnitude making this a reasonable definition). 
Second, we analyzed the behavior of the rate as a 
function of the entropy of the initial state (i.e., on the spread of 
the initial Wigner function). Again, we used the same definition of $t_c$ 
(for this study we considered a different set of parameters than the one
used to construct Figure 1 to assure that the chaotic sea can accommodate  
initial states with linear entropy up to ${\cal H}=4$). 
With this results we obtained the plots in Figure 3 where the linear 
dependence of $t_c$ on the initial entropy (mixed initial states produce 
entropy at a smaller rate) and its logarithmic dependence on the diffusion 
constant are shown.

%Entropy increase during decoherence is a consequence of the entangling 
%interactions between the system and its environment. 
%The master equation implies that that rate at which 
%the information leaks into the environment is given by $\dot{\cal H}=2 D
%\langle (\partial_p W)^2\rangle/\langle W^2\rangle$. Therefore, 
%the entropy production rate is closely related with the average 
%length-scale that is present in the Wigner distribution. Thus, the 
%transition we examined above  
%takes place at a time where the dominant length scales with diffusion as 
%$\sqrt D$ (as $\sigma_c$ does). This behavior cancels the diffusion dependence
% of $\dot{\cal H}$ which becomes entirely determined by the dynamics. 
Entropy production, a consequence of the entangling interactions between 
the system and its environment, plays a central role in studies of decoherence 
(for example, it is esential to find the ``pointer states'' of the 
system using the ``predictability sieve'' \cite{pointer}). Here, 
we presented results supporting the point of view stating \cite{zp95} 
that entropy production rate during decoherence could also be used as a 
diagnostic for quantum chaos. On the other hand these results 
also make evident the fact 
that, for quantum systems that are classically chaotic, the nature of the 
classical limit induced by decoherence is rather peculiar (and quite
different from the one corresponding to regular systems). Indeed, this limit 
exhibits an unavoidable source of unpredictability, being the rate at which 
information is lost into the environment  entirely fixed by the chaotic 
nature of the Hamiltonian of the system. 

This work was supported by grants from Anpcyt (PICT 01014), Ubacyt and 
Conicet.

%%%%%%%%%%%%%%%%%%%%%%%%%%%%%%%%%%%%%%%%%%%%%%%%%%%%%%%%%%%
%
%		References
%
%%%%%%%%%%%%%%%%%%%%%%%%%%%%%%%%%%%%%%%%%%%%%%%%%%%%%%%%%%%

%%%%%%%%%%%%%%%%%%%%%%%%%%%%%%%%%%%%%%%%%%%%%%%%%%%%%%%%%%%
%
%		Figure captions
%
%%%%%%%%%%%%%%%%%%%%%%%%%%%%%%%%%%%%%%%%%%%%%%%%%%%%%%%%%%%

\end{document}